\documentclass[a4paper,fleqn,usenatbib]{mnras}

\usepackage{mathptmx}

\usepackage[T1]{fontenc}
\usepackage{ae,aecompl}

\usepackage{graphicx}	
\usepackage{amsmath}	
\usepackage{amssymb}	
\usepackage{epstopdf}
\usepackage{textcomp}
\usepackage{multicol}
\usepackage{hyperref}

\title[]{Observations of V0332+53 during the 2015 Outburst using \textit{Fermi}/GBM, \textit{MAXI}, \textit{Swift}, and \textit{INTEGRAL}}

\author[Z. A. Baum et al.]{
Zachary A. Baum$^{1}$,
Michael L. Cherry$^{1}$,
James Rodi$^{2,3}$
\\
$^{1}$\textit{Department of Physics and Astronomy, Louisiana State University, Baton Rouge, Louisiana 70803, USA}\\
$^{2}$\textit{Universite\'\ de Toulouse; UPS-OMP; IRAP; Toulouse, France}\\
$^{3}$\textit{CNRS; IRAP; 9 Av. Colonel Roche, BP 44346, F-31028 Toulouse cedex 4, France}
}

\date{Accepted XXX. Received YYY; in original form ZZZ}

\pubyear{2016}

\begin{document}
\label{firstpage}
\pagerange{\pageref{firstpage}--\pageref{lastpage}}
\maketitle

\begin{abstract}
We present the lightcurves, spectra, and hardness-intensity diagram (HID) of the high mass X-ray binary V0332+53  using \textit{Fermi}/GBM, \textit{MAXI}, \textit{Swift}/BAT, and \textit{INTEGRAL} through its 2015 Type II outburst. We observe characteristic features in the X-ray emission (2-50 keV) due to periastron passages, the dynamical timescale of the accretion disc, and changes within the accretion column between a radiation-dominated flow and a flow dominated by Coulomb interactions. Based on the HID and the light curves, the critical luminosity is observed to decrease by $\sim5\%-7\%$ during the outburst, signaling a decrease in the magnetic field.
\end{abstract}



\section{Introduction}

Outbursts from the transient X-ray pulsar V0332+53 were detected in 1973 \citep{terrell1984}, 1983 \citep{makishima1990a}, 1989 \citep{makishima1990b}, 2004-2005 \citep{pottschmidt2005,mowlavi2006,lutovinov2015,caballero2016}, and recently a large Type II outburst starting in June 2015 
\citep{doroshenko2016,tsygankov2016,caballero2016,wijnands2016,cusumano2016}. These outbursts have typically been split into two categories, Type-I and Type-II outbursts. The relatively less luminous Type-I outbursts are more frequent, occurring near periastron passages and lasting a few days. Type-II outbursts have a peak luminosity near the Eddington luminosity  \citep{frank2002} and can last for several months. The peak X-ray intensity in the giant 1973 outburst reached $\sim$1.6 Crab in the 3-12 keV band \citep{terrell1984} while the intensity approached $\sim$1.2 Crab in the 1.3-12.2 keV band for the 2004-2005 outburst \citep{tsygankov2006}. The  Type-II outbursts in V0332+53 have typically had a duration of  2-3 months, with a decay time $\sim$15-20 days longer than the rise time, and have occurred approximately once every 10 years since the first detection in 1973. The optical companion, BQ Cam, is an O8-9 type main sequence star \citep{negueruela1999} which has been observed to brighten in optical and IR for several months prior to the X-ray flare \citep{goranskii2001,caballero2016}. The X-ray outbursts then occur due to the passage of the neutron star through the enhanced decretion disk of BQ Cam \citep{caballero2016}.

The X-ray spectrum of V0332+53 has usually been fit by a phenomenological model composed of a power law with exponential cutoff plus cyclotron absorption line \citep{makishima1990a} or an NPEX (Negative and Positive power-law with EXponential) model \citep{mihara1995} plus cyclotron absorption lines \citep{kreykenbohm2005, pottschmidt2005}. 
Based on RXTE observations of the 2004 event, \cite{pottschmidt2005} fit the spectrum with three cyclotron absorption lines at approximately 26, 52, and 74 keV, corresponding to a surface magnetic field of $2.7 \times 10^{12}$ G. Observations of the 2004 outburst  noted that the energy of the fundamental cyclotron line decreases with increasing luminosity and that the ratio of the energy of the first harmonic to that of the fundamental line is near 2 and increases with luminosity \citep{tsygankov2006,nakajima2010,lutovinov2015}. This has been interpreted as due either to changes in the height at which the cyclotron absorption line is produced \citep{becker2012} or to reflection from the neutron star surface \citep{poutanen2013}. \cite{nakajima2010} and \cite{tsygankov2010} also investigated possible hysteresis of the cyclotron absorption line energy, equivalent width, and pulsed fraction between the 2004-2005 outburst rise and decline, which might signal a change in the physics of the accretion column or emission. They found no sign of hysteresis in the 2004-2005 event. However, a decrease of the line energy has been observed over the 2015 outburst and attributed to a drop in the magnetic field during the outburst \citep{cusumano2016}.

\cite{reig2008} and \cite{reig2013} have created color-color diagrams (CCD) and hardness-intensity diagrams (HID) for a number of Be/X-ray binaries including V0332+53 during Type II outbursts.  Similar to LMXBs, the HIDs are characterized by two different accretion states with a "horizontal" branch corresponding to relatively low accretion rates and a "diagonal" branch corresponding to higher accretion. The transition between the horizontal and diagonal branches is interpreted to occur \citep{becker2012} at the point at which the luminosity reaches the critical luminosity (i.e., where the inward flow becomes radiation-dominated on the diagonal branch rather than governed by Coulomb interactions on the horizontal branch). 4U 0115+63 and V0332+53 were the only sources where hysteresis was observed in the HID during the time the source was on the diagonal branch, with the spectrum becoming softer during the outburst decline \citep{reig2008,reig2013}. The reported hysteresis in the HID for V0332+53 was observed for the 2004-2005 outburst \citep{reig2008}.

Between January 2012 and February 2015, the optical brightness of BQ Cam increased by $\sim$0.7 mag \citep{caballero2016}. On June 20 2015 (MJD 57193), an X-ray outburst was then reported to begin by \textit{MAXI} \citep{nakajima2015} followed by \textit{Swift} \citep{doroshenko2015}. The event rose to a peak of $\sim 1.7$ Crab in the 12-25 keV energy range after approximately 50 days and decreased back to quiescence after approximately 150 days. \cite{cusumano2016} have shown that the peak energy of the fundamental cyclotron line decreased from $29.2_{-0.3}^{+0.4}$ keV at the start of the event to $26.13_{-0.06}^{+0.07}$ keV after 56 days and then recovered to $27.68 \pm 0.15$ keV after 105 days, corresponding to a decrease in the surface magnetic field from $3.28 \times 10^{12}$ G to $2.93 \times 10^{12}$ G near the peak and then an increase back to $3.11 \times 10^{12}$ G at the end of the event (assuming a redshift of 0.3 corresponding to a $1.4$ M$_{\odot}$ neutron star with radius of 10 km). \textit{Fermi}/GBM observed the evolution of the pulsar spin frequency throughout the 2015 outburst\footnote{http://gammaray.msfc.nasa.gov/gbm/science/pulsars/ \linebreak lightcurves/v0332.html} and observed the spin-up of the pulsar from $\sim 228.505$ mHz at MJD 57200 to $\sim 228.525$ mHz at MJD 57270.

We have combined data from \textit{\textit{MAXI}}/GSC, \textit{Fermi}/GBM, \textit{Swift}/BAT, \textit{INTEGRAL}/JEM-X, and \textit{INTEGRAL}/SPI to study the light curves, spectral shape, and spectral evolution of V0332+53 during its 2015 Type II outburst. In Section~\ref{sec:IandO}, we briefly describe the instruments and the observations; in Section~\ref{sec:data}, we present the results of the analysis; and in Section~\ref{sec:disc}, we discuss the possible physical interpretations. In particular, we use the hardness-intensity diagram in combination with the light curve to provide information about the critical luminosity, and emphasize the role of periastron passages and changes in accretion state in determining the features of the light curve.

\section{Instruments and Observations}
\label{sec:IandO}
\subsection{\textit{MAXI}/GSC}
The \textit{MAXI} (Monitor of All Sky X-ray Image) Gas Slit Camera (GSC, \cite{matsuoka2009}) on the \textit{International Space Station (ISS)}, launched in August 2009, is a 5350 cm$^2$ array of Xe proportional counters operating at 2 - 30 keV. The instrument has a wide field of view (6 identical units with 160\textdegree $\times$ 1.5\textdegree\ FOV) and covers essentially the entire sky during each \textit{ISS} orbit (92 min). \textit{MAXI}/GSC observations in this work are averaged over each day, with $5\sigma$ source sensitivity $\sim 15$ mCrab. Observations are removed when the source is observed near the edge of the field of view of the GSC, where sensitivity drops. This behavior was confirmed using images from the \textit{MAXI} on-demand process\footnote{http://maxi.riken.jp/mxondem/}. \textit{MAXI} observed the type-II outburst of V0332+53 during 2015 from approximately periastron at the beginning of the rise through the peak and then over approximately 40 days on the decline, except for days when the source was out of or near the edge of the field of view.

\subsection{\textit{Fermi}/GBM}
The Gamma-ray Burst Monitor (GBM), one of the two instruments on board the \textit{Fermi Gamma-Ray Space Telescope}, observes steady and transient sources using earth occultation \citep{case2011, wilson2012}.  \textit{Fermi} was launched in June 2008 and consists of 14 detectors: 12 NaI detectors operating over the energy range 8 keV - 1 MeV and 2 bismuth germanate (BGO) detectors operating at 150 keV - 40 MeV. The NaI detectors are used for burst and pulsar analysis, providing time-tagged data with nominal 0.256 s time resolution and 8-channel spectral resolution. In the Earth occultation mode, typically 3-4 NaI detectors view an Earth occultation within 60\textdegree\ of the detector normal vector, allowing the instrument to observe $>85\%$ of the sky in a single orbit and the entire sky every $\sim26$ days. The two BGO detectors are located on opposite sides of the spacecraft and also view a large part of the sky in the high energy range. The \textit{Fermi}/GBM data presented here have been averaged over at least one day of observations, with one-day $3\sigma$ sensitivity being $\sim 100$ mCrab. Daily averages without a significant number of occultations have been removed, where the number of occultations required for a significant detection are dependent on the observed source flux. \textit{Fermi}/GBM observed the entire 2015 outburst with the exception of a few gaps where there were not a significant number of occultations of V0332+53.

\subsection{\textit{Swift}/BAT}
The \textit{Swift}/Burst Alert Telescope (BAT) is a 5200 cm$^2$ coded aperture telescope operating in the 14-195 keV range with a 2.0 steradian FOV, 17 arcminute resolution, and 1 - 3 arcminute location precision \citep{barthelmy2005}. Launched in Nov.\ 2004, BAT has been used to create a hard X-ray All-Sky Survey to a flux limit $\sim10^{-11}$ erg cm$^{-2}$s$^{-1}$ over 70 months \citep{baumgartner2013}. BAT accumulates detector plane maps approximately every five minutes in 8 energy bands. Sky coverage for transients is of $\sim 50 - 80$\% at $>20$ mCrab in one day \citep{tueller2010}. \textit{Swift}/BAT observed the entire 2015 outburst.

\subsection{\textit{INTEGRAL}}
The \textit{International Gamma-ray Astrophysics Laboratory} (\textit{INTEGRAL}) \citep{jensen2003} was launched in October 2002 with a highly eccentric 3-day orbital period.  With the X-ray monitor JEM-X \citep{lund2003}, the gamma-ray spectrometer SPI \citep{roques2003}, and the gamma-ray imager IBIS \citep{ubertini2003}, \textit{INTEGRAL} is able to study sources from 3 keV \(-\) 10 MeV.  For this work, data were analyzed from JEM-X because of its low energy coverage and SPI, the latter because of its better energy resolution relative to IBIS/ISGRI. During the V0332+53 outburst, \textit{INTEGRAL} observed the source mainly through two Target of Opportunity (ToO) periods.  The first spanned 2015 July 17 02:32:50 to July 18 03:03:15 UTC (MJD 57220-57221) during the rising part of the flare, and the second from 2015 July 30 11:57:51 to August 1 12:47:47 UTC (MJD 57233-57235) during the peak of the flare.

\section{Data Analysis}
\label{sec:data}

\subsection{Light Curves}

\begin{figure*}
  \centering 
    \includegraphics[scale=0.75, trim = 7mm 0mm 0mm 0mm, clip] {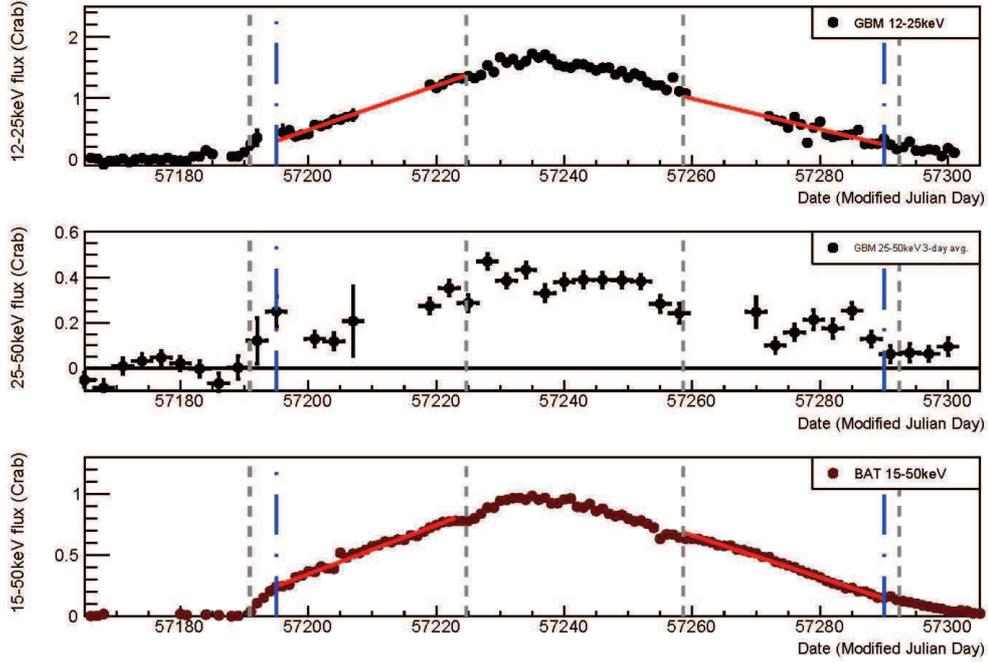}
  \caption{\textit{Fermi}/GBM and \textit{Swift}/BAT light curves. Vertical dashed lines mark times of periastron passages. Vertical dot-dashed lines mark transitions between horizontal and diagonal branches in hardness-intensity diagram.}
  \label{fig:hard_lc}
\end{figure*}

\begin{figure*}
  \centering
  \includegraphics[scale=0.75, trim = 7mm 0mm 0mm 0mm, clip]
{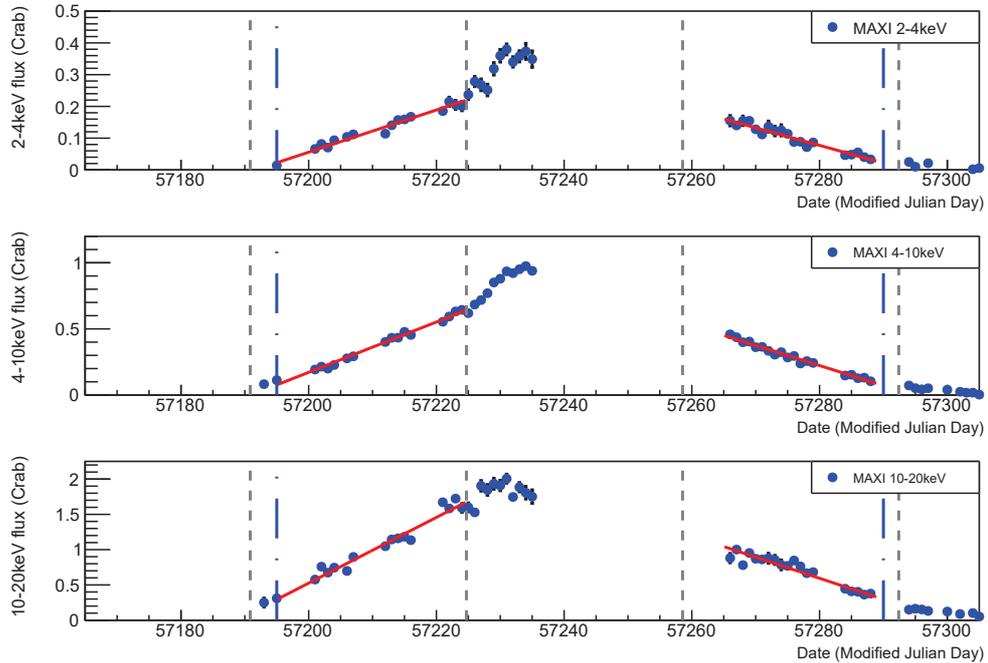} 
 \caption{\textit{MAXI}/GSC light curves. Vertical dashed lines mark times of periastron passages. Vertical dot-dashed lines mark transitions between horizontal and diagonal branches in hardness-intensity diagram.}\label{fig:MAXI_lc}
\end{figure*}

Figure~\ref{fig:hard_lc} shows the outburst light curves observed by \textit{Fermi}/GBM at 12-25 and 25-50 keV and Swift/BAT at 15-50 keV. Figure~\ref{fig:MAXI_lc} shows the \textit{MAXI}/GSC 2-4, 4-10, and 10-20 keV light curves of the 2015 outburst. The 2-4, 4-10, 10-20 keV \textit{MAXI}, 12-25 keV GBM, and 15-50 keV BAT light curves are made up of daily flux averages normalized to Crab units, while the 25-50 keV GBM light curve is made up of 3-day flux averages also normalized to Crab units.  Gaps in the GBM data are caused by not having a significant number of occultations on those days. Gaps in the \textit{MAXI} data are due to the source leaving the \textit{MAXI}/GSC field of view.

In both Figures~\ref{fig:hard_lc} and ~\ref{fig:MAXI_lc}, the times of periastron passage are marked by dashed-vertical lines \citep{doroshenko2016}. The outburst begins near or slightly before the periastron passage at MJD $57190.85$. There is an indication that the outburst begins slightly earlier at higher energies, although because all instruments are observing near their respective noise limits, the differences in sensitivity may play a role in the observed effect. If the outburst does indeed begin earlier in the hard X-rays, this could be related to the spectral hardening observed as V0332+53 transitions from low-accretion states \citep{tsygankov2016, wijnands2016}. In the hard X-rays, the flux then increases linearly until there is a "kink" at MJD $57195$. As discussed in Section \ref{ssec:disc_shape} below, this kink coincides with V0332+53 changing from the horizontal branch of its HID track to the diagonal branch and is marked with a vertical dotted/dashed line. The flux then increases linearly across all energy ranges until the second periastron passage at MJD $57224.7$, where an uptick in flux is seen in all light curves. During the outburst decline, BAT had the most complete coverage of the periastron passage at MJD $57258.55$, and there again appeared to be a change in the slope associated with the periastron passage. There is no noticeable kink in the flux as the source moved back to the horizontal branch of the HID at MJD $57290$ as was observed in the outburst rise. 

To test the hypothesis of \cite{mowlavi2006} which suggested an exponential decrease in flux throughout the outburst decline related to exponential emptying of the neutron star accretion disk, a straight line was fit to the \textit{Swift}/BAT data between the time of transition to the diagonal branch (radiation-dominated braking) at MJD $57195$ and $57223$, where the flux started to plateau just before periastron passage at MJD $57224.7$. The reduced-$\chi^2$ of this linear fit was $1.6$ ($27$ DOF). An exponential fit to the same time interval produced a reduced-$\chi^2$ of $6.7$ ($27$ degrees of freedom, DOF). A straight line was also fit to the data between the periastron passage at MJD $57258.55$ and the transition to the horizontal branch at MJD $57290$, with a reduced-$\chi^2$ of $1.7$ ($30$ DOF). An attempt to fit this interval of the outburst decline to an exponential decrease in flux was unsuccessful. The data are best fit by using a roughly linear increase and decline rather than an exponential increase or decline. In disagreement with \cite{mowlavi2006}, this suggests that the dominant factors in determining outburst shape are the times of periastron and transitions between radiation-dominated slowing (braking) of the inward flow and slowing due to Coulomb interactions.

In the BAT and GBM energy ranges, V0332+53 switches between modes of braking at a lower flux level during the outburst decline than during the rise. At the lower \textit{MAXI} energy range, however, no strong statements can be made due to a lack of observations near the beginning of the outburst. The decrease in flux between times of transition points to a more general drop in the luminosity at these times. This is, by definition, a drop in the critical luminosity, confirming the observed drop in magnetic field observed by \cite{cusumano2016}.

\subsection{Spectral Analysis}

\begin{figure*}
  \centering
  \includegraphics[scale=0.9, trim = 30mm 75mm 13mm 85mm, clip]
{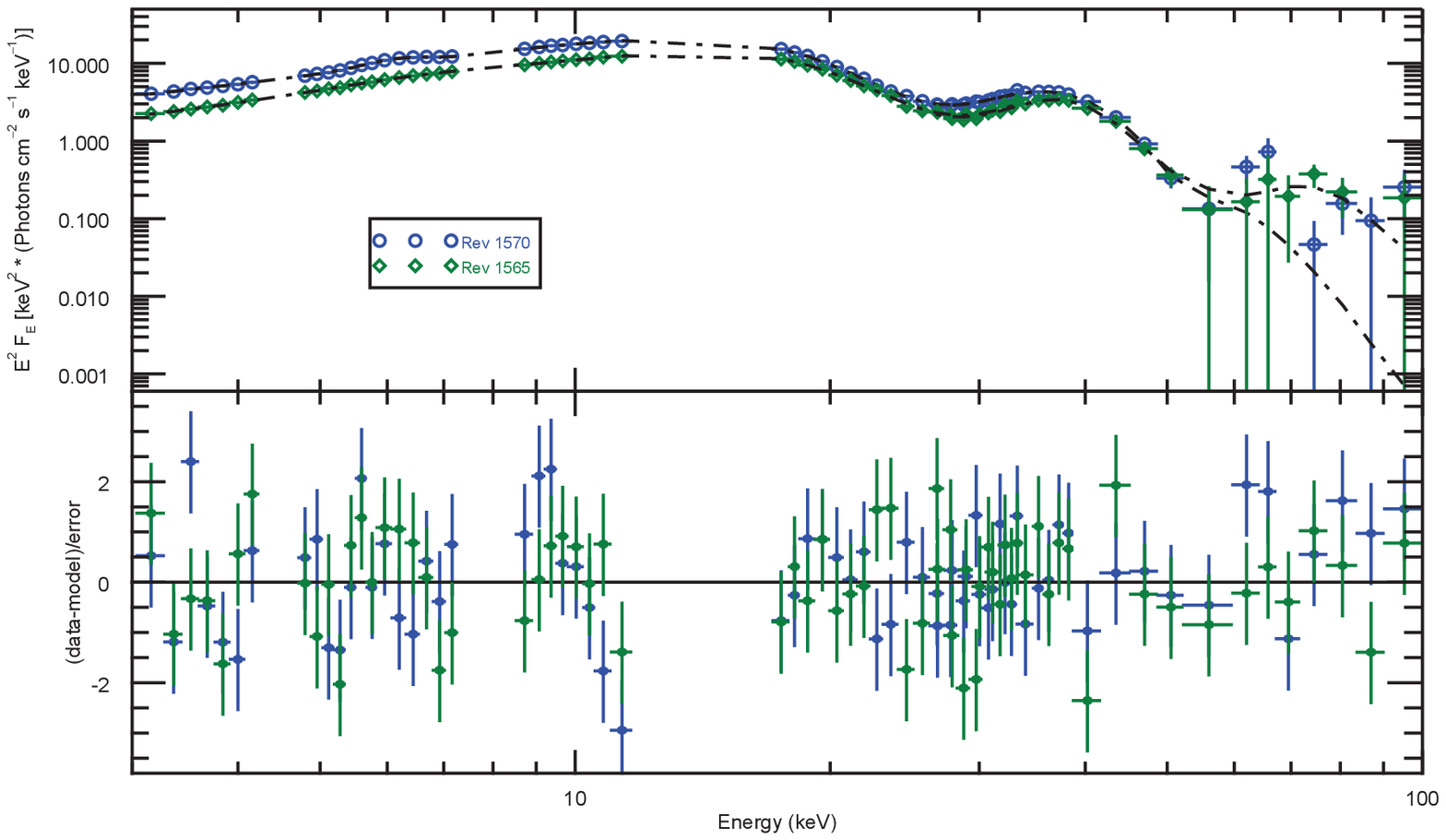}
  \caption{V0332+53 \textit{INTEGRAL} spectra during outburst rise (Rev 1565) and near peak (Rev 1570).}\label{fig:spec}
\end{figure*}

\begin{table*}
\begin{center}
\begin{tabular}{cccc}
\hline\hline
\textit{INTEGRAL} Rev. & 1565 & 1570 & \\
\hline
Start Time (SPI) & 57220.098 & 57233.490 & MJD\\
End Time (SPI) & 57221.175 & 57235.535 & MJD \\
Exposure Time (SPI) & $6.586\times10^{4}$ & $1.160\times10^{5}$ & s\\
Start Time (JEM-X) & 57220.106 & 57233.548 & MJD \\
End Time (JEM-X) & 57221.134 & 57235.535 & MJD \\
Exposure Time (JEM-X) & $6.479\times10^{4}$ & $1.313\times10^{5}$ & s \\
\hline
Cal. Factor & $1.06^{+0.04}_{-0.04}$ & $1.01^{0.02}_{0.02}$ & \\
kT & $6^{+0.5}_{-0.5}$ & $4.46^{+0.04}_{-0.04}$ & keV\\
Opt. Depth & $10.2^{+0.3}_{-0.2}$ & $10.30^{+0.06}_{-0.07}$ & \\
Norm. Const. & $0.55^{+0.05}_{-0.04}$ & $1.15^{+0.05}_{-0.05}$ & \\
E$_{cyc,1}$ & $27.3^{+0.2}_{-0.2}$ & $26.1^{+0.1}_{-0.1}$ & keV \\
$\sigma_{cyc,1}$ & $6.2^{+0.2}_{-0.2}$ & $5.4^{+0.1}_{-0.1}$ & keV \\
$\tau_{cyc,1}$ & $34^{+3}_{-4}$ & $22.0^{+0.8}_{-0.6}$ & \\
E$_{cyc,2}$ & $55^{+3}_{-2}$ & $53^{+6}_{-3}$ & keV \\
$\sigma_{cyc,2}$ & $11^{+2}_{-2}$ & $5^{+2}_{-1}$ & keV \\
$\tau_{cyc,2}$ & $72^{+36}_{-28}$ & $14^{+24}_{-6}$ & \\
E$_{Fe}$ & $6.1^{+0.2}_{-0.3}$ & $6.13^{+0.05}_{-0.04}$ & keV \\
$\sigma_{Fe}$ & $1.8^{+0.4}_{-0.3}$ & $0.5^{+0.1}_{-0.09}$ & keV \\
Norm. Const. & $0.14^{+0.06}_{-0.04}$ & $0.507^{+0.005}_{-0.004}$ & \\
Red. $\chi^2$ & 1.24 & 1.35 & \\
2-10 keV Flux & $1.22\times 10^{-8}$ & $2.02\times 10^{-8}$ & ergs/cm$^{2}$/s \\
\hline
\end{tabular}
\caption{Observation times and spectral fit results for the two \textit{INTEGRAL} observation periods. Spectral fit parameters used in the comptt, gabs, and gauss fit models are: Cal. Factor (scaling factor used to smoothly fit the JEM-X results to SPI), kT (electron plasma temperature), Opt. Depth (optical depth), and Norm. Const. (overall normalization) for the Comptonization spectrum; E$_{cyc,1,2}$, $\sigma_{cyc,1,2}$, and $\tau_{cyc,1,2}$ (peak energy, sigma, and optical depth for the first and second harmonic cyclotron lines); E$_{Fe}$, $\sigma_{Fe}$, and Norm. Const. (peak energy, sigma, and normalization for the iron line); overall reduced $\chi^2$; and the fitted 2-10 keV flux.}
\label{tab:spec}
\end{center}
\end{table*}

Figure~\ref{fig:spec} shows JEM-X and SPI spectra taken during Revolution 1565 (MJD 57220-57221), when the V0332+53 flux was increasing, and during Revolution 1570 (MJD 57233-57235), when V0332+53 was near the peak of its outburst. The spectra were fit in XSPEC v.$12.9$ \citep{arnaud1996} using an optically thick Comptonization model (compTT in XSPEC) while the cyclotron resonant features were fit using gaussian absorption profiles (gabs in XSPEC). We note that other authors have used a powerlaw with a high-energy cutoff (cutoffpl in XSPEC) to fit the spectra of HMXBs \citep{mowlavi2006,tsygankov2006,cusumano2016}. Here we have used a physically motivated optically thick comptonization model which gave a similar goodness of fit to the powerlaw with a high-energy cutoff when fitting. A gaussian component was also needed to describe the iron line, which was more significant in the Revolution 1565 spectrum than in the Revolution 1570 spectrum. JEM-X channels 8-10, 24-27, and 36-45 were ignored in the fitting because of instrumental features located in those channels that could not be fitted appropriately. The seed photon temperature was fixed at 0.5 keV following \cite{tsygankov2016}, where a blackbody spectrum was observed with a temperature of 0.5 keV in the propeller regime (after accretion had stopped) for the 2015 outburst. The best fit results are shown in Table~\ref{tab:spec}. 

In Table~\ref{tab:spec}, an anti-correlation is observed between luminosity and the cyclotron line energies as has been reported for the 2005 outburst \citep{mowlavi2006, tsygankov2006} and previously for this 2015 outburst \citep{cusumano2016}. The observed fundamental line energies reported in Table~\ref{tab:spec} agree well with those reported in \cite{cusumano2016} for spectra taken at similar times. Differences in the fundamental line width and first harmonic line energy may be due to differences between the \textit{Swift} and \textit{INTEGRAL} instrument responses or the slightly different continuum models used (cutoffpl in \cite{cusumano2016} and compTT in this work). 

\subsection{Color Analysis}

\begin{figure*}
  \centering
  \includegraphics[scale=0.75, trim = 5mm 0mm 0mm 0mm, clip] {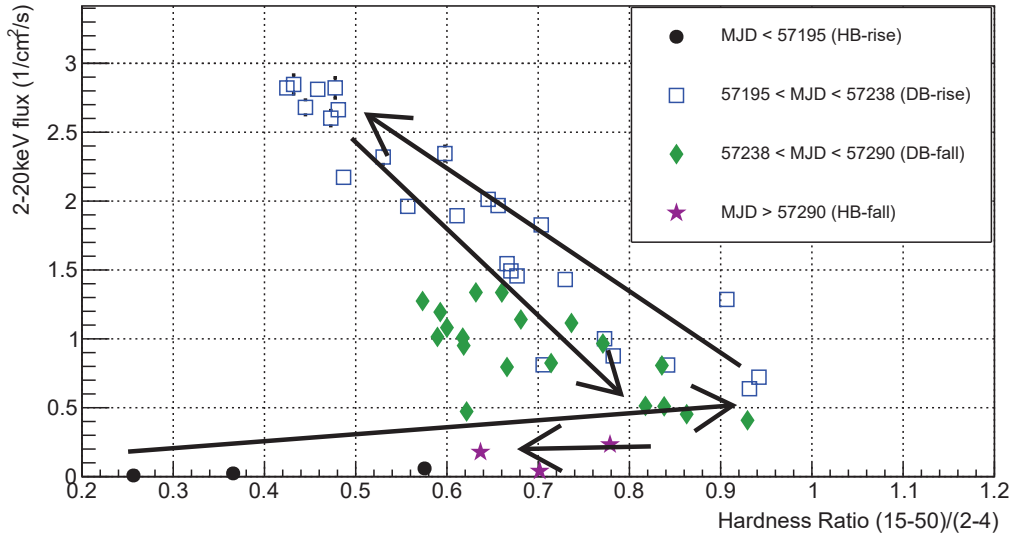}
 \caption{2-20 keV intensity vs (15-50 keV)/(2-4keV) hardness ratio measured during the 2015 outburst. Arrows show the evolution of the source as flux increases through the rising horizontal and diagonal branches (HB-rise and DB-rise) and as flux decreases through the falling diagonal and horizontal branches (DB-fall and HB-fall). Error bars are included for both flux and hardness ratio, but are smaller than marker size except at the highest fluxes.}\label{fig:hid}
\end{figure*}

X-ray colors, or hardness ratios, provide a means to quantify the spectral shape during the outburst and are shown in the HID in Figure~\ref{fig:hid}. Fluxes from \textit{Swift}/BAT and \textit{MAXI}/GSC were used to create the (15-50keV)/(2-4keV) hardness ratios during the 2015 outburst.  In Figure~\ref{fig:hid}, the source can be seen to move from the rising horizontal branch (becoming harder at low intensity), to the rising diagonal branch (becoming softer with increasing luminosity), and then down the falling diagonal branch (becoming harder with decreasing luminosity) and finally along the falling horizontal branch (becoming softer at low intensity) as the outburst comes to an end.

The transition from the horizontal branch to the diagonal branch is expected \citep{becker2012} to be the point at which the luminosity reaches the critical luminosity (i.e., where the inward flow becomes radiation-dominated on the diagonal branch rather than governed by Coulomb interactions on the horizontal branch). This occurs at approximately MJD 57195 in the HID, corresponding to the change in slope of the light curve at the same time. The peak in the light curve (near MJD 57238) corresponds to the transition from the rising diagonal to falling diagonal branch. The transition from falling diagonal to falling horizontal branch occurs at approximately MJD 57290, where Figures ~\ref{fig:hard_lc} and ~\ref{fig:MAXI_lc} show slight although not conclusive indications of flattening of the light curves.

Figure~\ref{fig:hid} shows a difference in color between the rise and decay of the outburst (i.e., hysteresis in the HID), with the source being harder during the outburst rise than during the fall. Despite the incomplete observations of the outburst decline, the hysteretic behavior observed is much larger than the plotted statistical errors and cannot be explained by the natural spread of the data.

\section{Discussion}
\label{sec:disc}

\subsection{Outburst Shape}
\label{ssec:disc_shape}
Using the most recent orbital parameters from \cite{doroshenko2016}, specifically the times of periastron, the orbital modulation of the light curve becomes apparent. The outburst begins in the 12-25 keV band just prior to periastron at MJD $57190.85$. In the GBM and BAT light curves, there appears to be a flattening of the light curve just before the second periastron at MJD $57224.7$ followed by a steepening. During the decline, the BAT light curve shows a flattening around the third periastron at MJD $57258.55$. The flux is too low at the fourth periastron to allow any definitive statement about the light curve behavior. 

During the outburst rise, the difference of a few days between periastron and the increases in flux can be roughly described by the accreted material from the decretion disk of BQ Cam traveling through the accretion disk of the neutron star and reaching the X-ray emission regions. For a separation between BQ Cam and the neutron star at periastron of $\sim3.6\times10^{7}$ km, with a dynamical timescale determined by the neutron star mass of $1.4$ M$_{\odot}$, a delay of $\sim6$ days is expected \citep{doroshenko2016,negueruela1999}. It is likely however that increases in the accretion rate would begin before the neutron star reaches periastron. It should also be noted that the accretion disk would not extend all of the way to the neutron star surface due to the large magnetic field of the neutron star pulling material from the accretion disk at the Alfven radius. In this case, the energy dependence of the outburst light curve would then be determined by the details of how the accretion rate affects the soft and hard X-ray emission regions/processes. Based on the dependence of the hard X-ray light curve on the mode of braking within the accretion column, we are led to conclude that at least the X-rays above $\sim12$ keV (lower limit of \textit{Fermi}/GBM energy band) must originate from within the accretion column to create such a dependence \citep{becker2012}. No definitive conclusions based on Figure~\ref{fig:MAXI_lc} can be made about the emission region of the softer X-rays due to the scarcity of \textit{MAXI} observations near the time of transition across the critical luminosity.

\cite{mowlavi2006} have suggested that the 3-60 keV \textit{INTEGRAL} light curve during the decline of the 2005 outburst could be modeled by an exponential decay with a folding time of 20-30 days before becoming linear near the end of the outburst. They argue that the suggested behavior is similar to that in LMXBs, where a hot disk illuminated by the central star empties at a rate proportional to its mass, producing an exponential decay of the emitted flux. As in \cite{mowlavi2006}, it follows that when the disk cools and is no longer uniformly heated, the flux then follows a linear decrease. 

If the exponential behavior was real as suggested in \cite{mowlavi2006}, then one would expect to see this throughout the entire outburst decay. Figures ~\ref{fig:hard_lc} and ~\ref{fig:MAXI_lc} show a roughly linear decay between periastron at MJD $57258.55$ and the transition between accretion states at MJD 57290 \textit{which does not support the description of exponentially decreasing flux throughout the outburst decay}. However, there is  the possibility that the dips observed just before periastron in the GBM and BAT light curves could still be described by an exponentially emptying accretion disk. This is supported by the outburst decline having more modulation on top of the linear decay when compared to the linear increase during the outburst rise, because the accretion disk would be more likely to only be filled partially during the decline.

\subsection{Critical Luminosity}

In Figure~\ref{fig:hard_lc}, the flux level of V0332+53 as it transitions across the critical luminosity is lower during the outburst decline than during the outburst rise. This is a signal that the critical luminosity has dropped, and by proxy, the magnetic field has also decreased. This confirms the result of \cite{cusumano2016} that based on the decrease of the cyclotron resonance line energy the magnetic field dropped during the outburst. Combining our knowledge of the times where V0332+53 crosses the critical luminosity with the spectra reported by \cite{cusumano2016}, we can evaluate the drop in the critical luminosity. Interpolating linearly between the spectra taken by \textit{Swift} on MJD 57193 and 57201 we estimate a critical luminosity of $\sim4.4\times10^{37}$ erg/s on 57195. Similarly, we estimate a critical luminosity of $\sim4.1\times10^{37}$ erg/s on MJD 57290 using the \textit{Swift} spectra taken on MJD 57277 and 57293, corresponding to a $\sim7\%$ decrease. This can be compared to the results obtained by using the fundamental cyclotron line energy to estimate the magnetic field and then the critical luminosity \citep{becker2012}. A fundamental line energy of $\sim29.0$ keV for MJD 57195, interpolating from \cite{cusumano2016}, corresponds to a surface magnetic field of $\sim2.5\times10^{12}$ G and a critical luminosity of $\sim4\times10^{37}$ erg/s. For MJD 57290 we interpolate a fundamental line energy of $\sim27.6$ keV, magnetic field of $\sim2.38\times10^{12}$ G, and a critical luminosity of $\sim3.79\times10^{37}$ erg/s, corresponding to a $\sim5\%$ drop in the critical luminosity. We note that no correction has been made here for gravitational redshift effects.

\cite{doroshenko2016b} have noted that the spin-up rate is higher for a given luminosity during the rising phase than during the decline. They argue that this is the opposite of what is expected from the connection between the torque exerted on the neutron star and the accretion flow, and suggest as a result that the decrease in the cyclotron line energy may be due to a change in the size of the magnetosphere rather than to a decrease in the magnetic field. On the contrary, the consistency shown above between the critical luminosity based on the times derived from the light curves and the HID, and the critical luminosity determined from the line energies seems to support the \cite{cusumano2016} suggestion of a direct connection between the cyclotron line energies and a decreasing magnetic field.

\subsection{Hysteresis in the HID}

Hysteresis in the HID can be seen in Figure~\ref{fig:hid}, where the falling diagonal branch is softer than the rising diagonal branch. Similar behavior has been observed during the 2004-2005 event \citep{reig2008,reig2013} despite the poor coverage of the outburst rise. This does not appear to be related to the decrease in magnetic field during the event suggested by the observed decrease in cyclotron line energy in 2015, since hysteresis was observed in the HID with no observed change in the magnetic field between the rise and decay of the 2004-2005 event \citep{reig2008,nakajima2010}. The lack of a connection between the magnetic field and hysteresis in the HID suggests that the cause of the hysteresis is likely located outside of the central accretion column.

In our case, we see a difference between the spectral fits for \textit{INTEGRAL} Revolutions 1565 and 1570 in the electron plasma temperature, which cools from $\sim6$ keV to $\sim4.5$ keV near the outburst peak. More efficient cooling of the electron plasma later in the outburst could be a possible explanation for the hysteresis observed in Figure~\ref{fig:hid}, where the hard X-rays more efficiently cool the plasma as accretion increases. The hard X-rays would be down-scattered during this process, resulting in a softer spectrum as the outburst rises to maximum.

\section{Conclusions}
We have analyzed data from \textit{MAXI}/GSC, \textit{Fermi}/GBM, \textit{Swift}/BAT, \textit{INTEGRAL}/JEM-X, and \textit{INTEGRAL}/SPI taken during the Type II outburst of the Be/X-ray binary system V0332+53 in 2015. The complex features in the light curve are correlated with the times of periastron passage, changes in the mode of braking within the accretion column, and the dynamical timescales of the accretion disk. In agreement with previous measurements, we see an anticorrelation of the fundamental cyclotron line energy with the luminosity. We also find that the X-ray spectra are well described by an optically thick comptonization model where the electron plasma is more efficiently cooled through Compton down-scattering as the outburst progresses. We track V0332+53 along its path on a hardness-intensity diagram throughout the outburst, observing clear indications of hysteresis (i.e., a softening of the spectrum) during the outburst. This behavior is similar to that observed in V0332+53 in its 2004-2005 outburst and in 4U 0115+63 \citep{reig2013}. Combining our knowledge of the times of transition between accretion states with the \textit{Swift} spectra \citep{cusumano2016}, we estimate that the critical luminosity decreases by $\sim 7\%$ from the rise of the outburst to the decline. 

\section{Acknowledgements}
This paper includes data collected by the \textit{Fermi} and \textit{Swift} missions, funded by the NASA Science Mission directorate. We especially appreciate the support from Colleen Wilson-Hodge and the other members of the GBM Earth Occultation team. MAXI data were provided by RIKEN, JAXA, and the MAXI team. Analysis of INTEGRAL data was supported by ESA and NASA through CNES. ZAB appreciates support from NASA EPSCoR and the Louisiana Board of Regents.







\bsp	
\label{lastpage}
\end{document}